\documentclass[reprint,onecolumn,superscriptaddress,amsmath,amssymb,aps,prl,floatfix]{revtex4-2}
\usepackage{graphicx}
\usepackage{upgreek}
\usepackage{bm}
\usepackage{hyperref}
\usepackage{todonotes}
\usepackage{nicefrac}
\usepackage[normalem]{ulem}

\begin{document}
	
	\title{Observation of an Anomalous Hall Effect in Single-crystal Mn\textsubscript{3}Pt}
	
	\author{Bel\'{e}n E. Zuniga-Cespedes}
	\affiliation{Max Planck Institute for Chemical Physics of Solids, D-01187 Dresden, Germany}
	\author{Kaustuv Manna}
	\affiliation{Max Planck Institute for Chemical Physics of Solids, D-01187 Dresden, Germany}
	\affiliation{Department of Physics, Indian Institute of Technology Delhi, Hauz Khas, New Delhi 110016, India}
	\author{Hilary M. L. Noad}
	\affiliation{Max Planck Institute for Chemical Physics of Solids, D-01187 Dresden, Germany}
	\author{Po-Ya Yang}
	\affiliation{Max Planck Institute for Chemical Physics of Solids, D-01187 Dresden, Germany}
	\author{Michael Nicklas}
	\affiliation{Max Planck Institute for Chemical Physics of Solids, D-01187 Dresden, Germany}
	\author{Claudia Felser}
	\affiliation{Max Planck Institute for Chemical Physics of Solids, D-01187 Dresden, Germany}
	\author{Andrew P. Mackenzie}
	\affiliation{Max Planck Institute for Chemical Physics of Solids, D-01187 Dresden, Germany}
	\affiliation{Scottish Universities Physics Alliance (SUPA), School of Physics and Astronomy, University of St. Andrews, St. Andrews KY16 9SS, United Kingdom}
	\author{Clifford W. Hicks}
	\affiliation{Max Planck Institute for Chemical Physics of Solids, D-01187 Dresden, Germany}
	\affiliation{School of Physics and Astronomy, University of Birmingham, Birmingham B15 2TT, United Kingdom}
	
	\begin{abstract}
		The Mn$_3$X family of compounds was the first in which a large anomalous Hall effect (AHE) was predicted to arise from a purely antiferromagnetic structure, due to the Berry
		curvature in momentum space.Nearly simultaneously with this prediction, a large AHE was observed experimentally in one of the hexagonal members of this family, Mn$_3$Sn. Aligning antiferromagnetic domains, a necessary step for observation of the AHE, is more challenging for the cubic members of the Mn$_3$X family, due to a combination of a smaller
		spontaneous ferromagnetic moment and much stronger magnetic anisotropy. Here, we use a combination of uniaxial stress and applied magnetic field to align domains of bulk
		single-crystal Mn$_3$Pt, and demonstrate for the first time a substantial AHE in a bulk sample of a cubic member of the Mn$_3$X family. The AHE remains locked in with essentially no quantitative variation when the stress is ramped back to
		zero, which shows that it is not a consequence of any stress-induced ferromagnetic moment.
	\end{abstract}
	
	\maketitle
	
	\section{Introduction}
	
	The anomalous Hall effect (AHE) was long thought to be a direct consequence of the magnetisation $M$ generated by ferromagnets~\cite{Pugh32_PR}.  However, in studies of the ferromagnet SrRuO$_3$ the anomalous Hall conductivity $\sigma_\text{xy}$ was found to have a nonmonotonic dependence on $M$~\cite{Fang03_Science}, and instead was found to be
	explainable as an integral of the spin Berry curvature in momentum space, following the proposal of Karplus and Luttinger~\cite{Karplus54_PR}. This Berry phase mechanism is now
	understood to be the dominant mechanism for the intrinsic anomalous Hall effect~\cite{Nagaosa10_RMP}. One feature of this mechanism is that even antiferromagnets can generate a
	large AHE, as predicted, for example, for antiferromagnetism on a distorted fcc lattice~\cite{Shindou01_PRL}, and observed experimentally in
	Pr$_2$Ir$_2$O$_7$~\cite{Machida10_Nature}. A large AHE is also predicted for the Mn$_3$X family of compounds~\cite{Chen14_PRL, Kuebler14_EPL}, which are also antiferromagnetic. It
	has been demonstrated experimentally in bulk, single-crystal Mn$_3$Sn~\cite{Nakatsuji15_Nature, Li19_NatComm}, Mn$_3$Ge~\cite{Nayak16_SciAdvances, Kiyohara16_PRA}, and
	Mn$_3$Ga~\cite{Liu17_SciReports}. A large, antiferromagnetically-induced AHE is a potentially valuable discovery for memory applications~\cite{Otani21_APL,
		NakatsujiReview}. 
	
\begin{figure}[ptb]
		\centering
		\includegraphics[width=121mm]{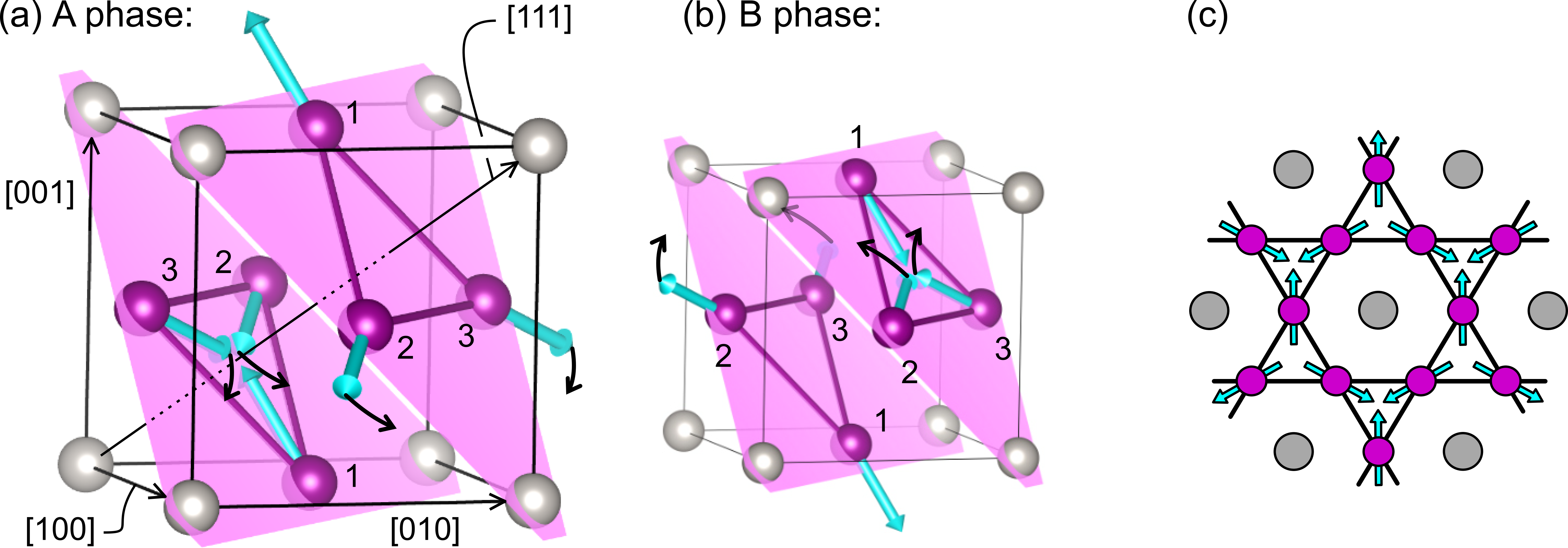}
		\caption{\label{fig:magStructure} Unit cell of Mn$_3$Pt, showing the magnetic structure for $T < T_\text{t} \approx 400$~K. The Mn sites are numbered for reference. The curved arrows show the expected spin rotation when the lattice is compressed along the $[001]$ axis, under an assumption of in-plane spin rotation. (a) Illustration of the A phase with the (111) spin plane. (b) Illustration of the B phase with the (111) spin plane. (c) A view of the magnetic structure along the [111] axis.}
	\end{figure}

	These three compounds are all hexagonal. Other members of the Mn$_3$X family --- Mn$_3$Pt, Mn$_3$Ir, and Mn$_3$Rh --- are cubic. The cubic compounds are also expected to have large
	anomalous Hall effects, but so far this has not been demonstrated in bulk single crystals. To observe the AHE in these compounds the magnetic structure must be trained, meaning that a subset of the
	possible antiferromagnetic domains must be selected, so that the AHE generated by individual domains does not cancel. It is easier to do so for the hexagonal compounds because they
	have an inverse-triangular spin configuration~\cite{Nagamiya82_SSC}, within which the spins do not point in crystallographically equivalent directions. One consequence is that spin canting results in a net ferromagnetic moment, 3~m$\upmu_\text{B}$/Mn in Mn$_3$Sn and 7~m$\upmu_\text{B}$/Mn in Mn$_3$Ge, upon which applied magnetic
	field can act to select domains. Another is that the coercive field for domain re-orientation is low. The cubic compounds, on the other hand, have a triangular spin
	configuration in which the spins do point in equivalent directions. Therefore, the magnetic anisotropy is 2--3 orders of magnitude higher than for the hexagonal
	compounds~\cite{Szunyogh09_PRB, Duan15_APL, Kota08_IEEE, Umetsu06_APL}, a property that has made one member of this family, Mn$_3$Ir, technologically important as an exchange-bias
	layer~\cite{tsunodaL12PhaseFormation2006,tsunodaExchangeAnisotropyL12006,aleyTextureEffectsIrMn2008,kohn2013antiferromagnetic}. The room-temperature magnetic structure of Mn$_3$Pt
	is shown in Fig.~\ref{fig:magStructure}. The spins lie essentially within $\{ 111 \}$ planes, within which the Mn atoms are arranged in a kagom\'{e} lattice. Cubic anisotropy is
	expected to introduce a slight rotation of the spins out of the plane~\cite{Szunyogh09_PRB, LeBlanc13_PRB}, resulting in a small net ferromagnetic moment.
	
	A polarisable AHE has been observed in thin films of Mn$_3$Pt~\cite{Liu18_NatureElectronics, Mukherjee21_PRM}, Mn$_3$Ir~\cite{Iwaki20_APL}, and also Mn$_3$NiN~\cite{Boldrin19_PRM},
	which is also cubic. The polarisability of thin films may be a consequence of the combination of epitaxial strain and piezomagnetism: under tetragonal lattice distortion, the
	spin directions become inequivalent, potentially causing spin canting and inducing a net ferromagnetic moment on which the field can act. Stress-driven spin canting has been
	observed for Mn$_3$Sn~\cite{Ikhlas22_NatPhys}. For thin films of Mn$_3$Pt and Mn$_3$Ir, the observed AHE was of the same order as theoretical predictions developed for bulk
	lattices~\cite{Zhang17_PRB}. However, observations might be substantially affected by interface effects and disorder, and for this reason it is important to observe the AHE in
	bulk single crystals.
	
	Here, we report a large, hysteretic AHE in bulk single-crystal Mn$_3$Pt, using a combination of applied field and uniaxial stress to train the magnetic structure. We study
	Mn$_3$Pt rather than Mn$_3$Ir because the ductile nature of Mn$_3$Ir makes it difficult to apply uniaxial stress. The AHE appears as a stress-induced field hysteresis in the Hall
	effect. The AHE remains locked in when the stress is released, which shows that stress-induced magnetism is not itself the source of the AHE, in agreement with the
		expectation that it is the antiferromagnetic structure that generates the AHE.

	\section{Methods}
	
\begin{figure}[htb]
		\centering
		\includegraphics[width=61mm]{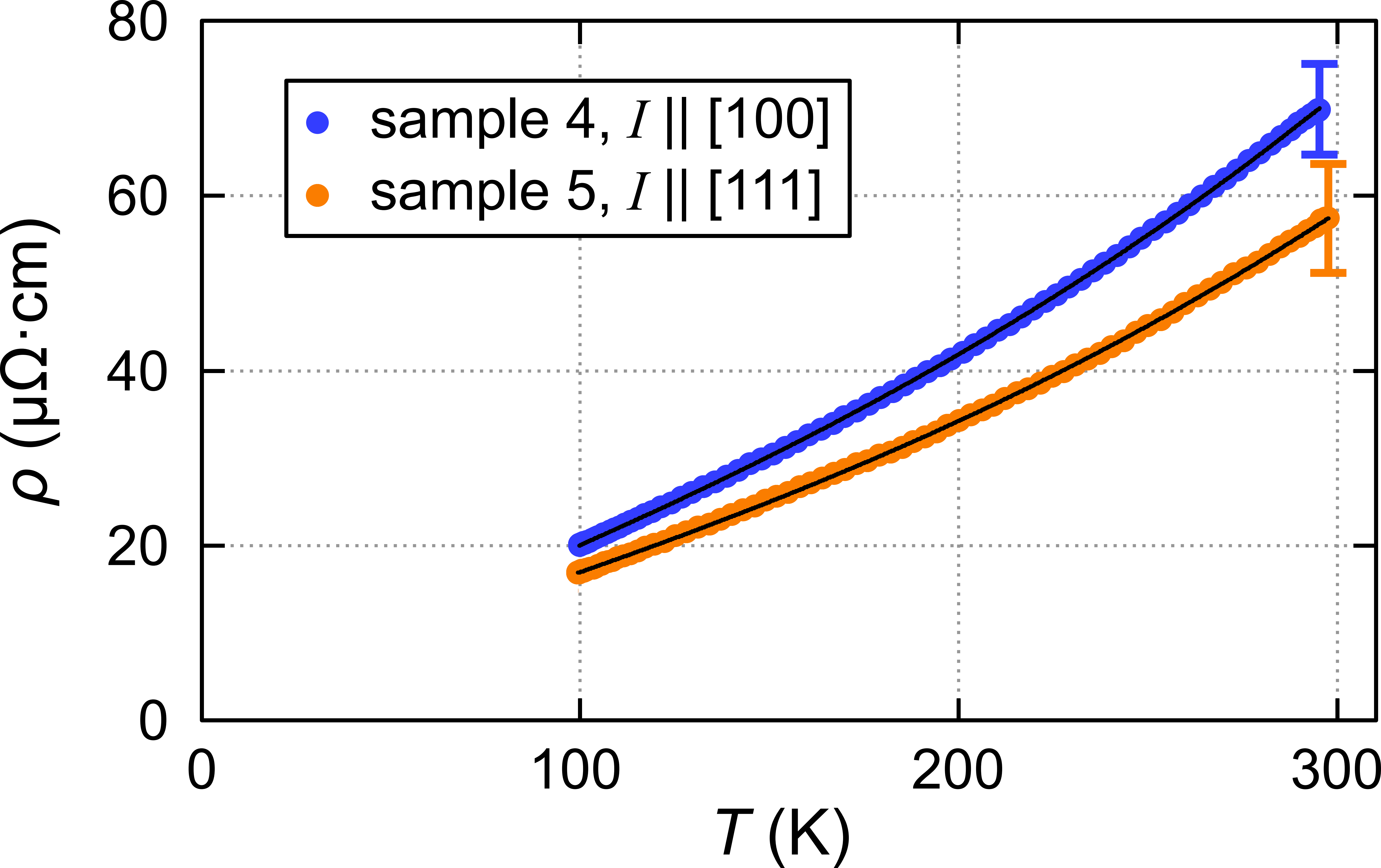}
		\caption{\label{fig:resistivity} Temperature dependence of the resistivity of two samples of Mn$_3$Pt, at zero stress. The black lines are third-order polynomial fits, used later
			to calculate the Hall conductivity. The error bars indicate systematic uncertainty due to error on the placement of contacts.}
	\end{figure}

	Single crystals of L1$_2$-Mn$_3$Pt were grown by the Bridgeman-Stockbarger method~\cite{Li19_MaterialsTodayPhysics}. All
	samples studied here were cut from the same original crystal.  Resistivity data from two samples are shown in Fig.~\ref{fig:resistivity}. The resistivity at 300~K is $65 \pm
	3$~$\upmu\Omega\cdot$cm, far smaller than the 175~$\upmu\Omega\cdot$cm measured for the epitaxial films of Ref.~\cite{Liu18_NatureElectronics}. 
	
\begin{figure*}[ptb]
		\centering
		\includegraphics[width=150mm]{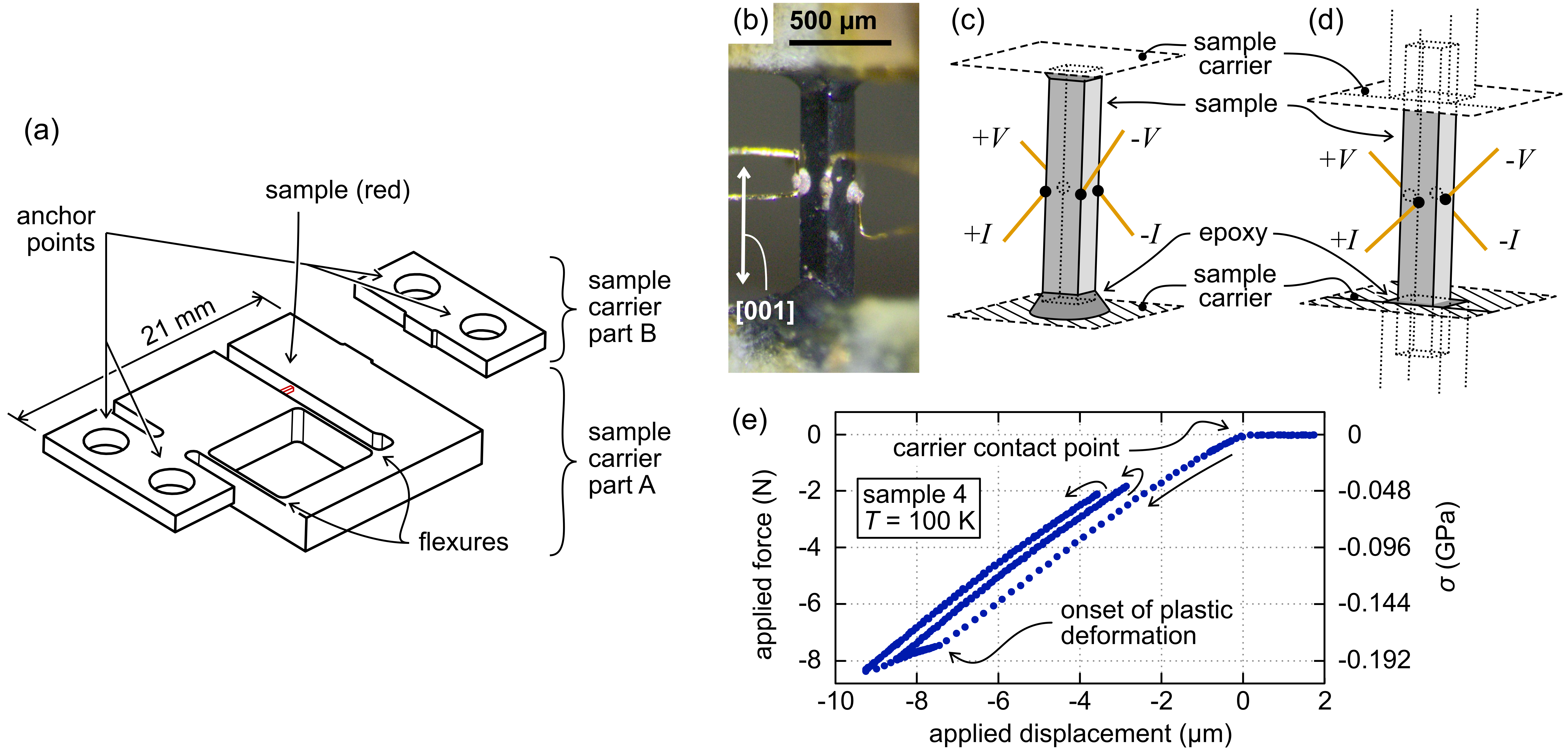}
		\caption{\label{fig:setup} Experimental setup. (a) Illustration of the sample carrier used for samples 1 and 4. The carrier is mounted onto a piezoelectric-driven cell that controls the
			distance between the pairs of anchor points. (b) Photograph of sample 1. (c) Schematic of the configuration for samples 1 and 3. The sample is held with epoxy between surfaces
			of the sample carrier, such that force is applied through the end faces of the sample. (d) Schematic of the configuration for sample 2. The ends of the sample are inserted
			into guides and secured with epoxy, such that force is applied to the sample dominantly through the side faces. The mounting epoxy for all samples was Stycast 2850FT, cured at
			$65^\circ$~C for 180~minutes. (e) Applied force versus applied displacement for sample 4 at 100 K.  The point labelled ``carrier contact point'' is where the two parts of
			the carrier come into contact.  The onset of plastic deformation is also indicated.  The stress scale, on the right, is the stress applied to the sample.}
	\end{figure*}

	Stress was applied using a piezoelectric-based uniaxial stress cell that incorporates sensors of both the force and displacement applied to the sample~\cite{Barber19_RSI}. To allow
	in situ calibration of the zero-stress point, samples were mounted onto carriers that allowed application of compressive but not tensile stress. Three different, though conceptually
	equivalent, designs were used; one is shown in Fig.~\ref{fig:setup} (a). It comprises two parts, and force is applied to the sample when they are brought together. 
	
	Samples 1, 2, and 3 had a longitudinal configuration: compression along $[001]$ and $\mathbf{H} \parallel [001]$. Qualitatively, this configuration matches that of the thin
	film study of Ref.~\cite{Liu18_NatureElectronics}, in which the epitaxial strain yielded $a > c$, where $a$ is the lattice constant along $[100]$ and $[010]$.  This configuration
	obliges us to adopt a non-optimal contact configuration for measurement of the Hall resistivity: as shown in Fig.~\ref{fig:setup} (b--d), current must be applied across the sample
	width. In a conventional configuration, the Hall voltage is:
	\begin{equation}
		\label{eq:HallVoltage}
		V_\text{y} = \rho_\text{xy} I_\text{x} / t.
	\end{equation}
	$t$ is the sample thickness. $\rho_\text{xy} = R_\text{H}B + \rho_\text{AHE} + \rho_\text{offset}$, where $R_\text{H}$ is the Hall coefficient, $\rho_\text{AHE}$ the anomalous Hall
	resistivity, and $\rho_\text{offset}$ is an offset due to misalignment of contacts. For samples 1--3, we applied Eq.~\ref{eq:HallVoltage} to obtain an effective thickness
	$t_\text{eff}$, so that $R_\text{H}$ at zero stress would match that measured in sample 4, which had a conventional Hall bar configuration. This $t_\text{eff}$ was then used
	to obtain the stress-driven changes in $\rho_\text{xy}$. Due to uncertainty in the true contact configuration, it is unlikely that more sophisticated analysis would yield better
	accuracy.
	
	Sample 4 had a transverse configuration, with stress along $[100]$ and $\mathbf{H} \parallel [001]$, allowing a conventional contact configuration. A normal Hall coefficient of
	$R_\text{H} = -0.057$~$\upmu\Omega\cdot$cm/T was observed at 298~K. For comparison, $R_\text{H}$ of a 20~nm-thick film of Mn$_3$Pt in
	Ref.~\cite{Liu18_NatureElectronics} was $-0.044$~$\upmu\Omega\cdot$cm/T at 365~K.
		
	All stress values quoted in this paper are based on a room-temperature calibration of the force sensor of the stress cell, and are accurate to within 10\%. The flexures of the
	sample carriers constitute springs placed in parallel with the samples; their total spring constant is 0.044~N/$\upmu$m for the carrier illustrated in Fig.~\ref{fig:setup} (a), and
	0.023~N/$\upmu$m and 0.020~N/$\upmu$m for the other two designs. These spring constants are at most 5\% of the spring constant of the sample system (see Table I), meaning that at
	least 95\% of the applied force went into the sample rather than the carrier flexures.
 
	Force-displacement data for sample~4 are shown in Fig.~\ref{fig:setup} (e). There are two prominent features. One is the point where the two parts of the carrier come into contact, allowing force to be transmitted to the sample. 
	Another is plastic deformation: the force-displacement relation transitioned to a shallower slope when the stress in this sample became
	larger than $-0.18$~GPa. (Negative values denote compression.) As shown in Table I, plastic deformation began at a similar stress for all samples where it was observed, even at
	very different temperatures, and even for different mounting configurations (e.g.: Fig. \ref{fig:setup} (c--d)) in which the stress field within the epoxy that holds the sample
	would have differed. We interpret this consistency, and the fact that with the same mounting epoxy far higher stresses have been achieved in other materials ~\cite{Barber19_RSI},
	as showing that the plastic deformation occured in the sample rather than the mounting epoxy. Evidently, the elastic limit of Mn$_3$Pt is much lower than that of Mn$_3$Sn, where
	elastic stresses larger than 1~GPa could be applied at temperatures near room temperature~\cite{Shirer20_APL}. The experiment here was a constant battle with plastic deformation.
	We only report data taken in the elastic regime. Within the elastic limit and the maximum field in our measurement system, 12~T, we obtained clear magnetic hysteresis, but not
	complete polarisation of the domains. 
	
\begin{table}[b]
	\caption{Key sample parameters. $k$ is the spring constant of the sample system, which is the sample itself in series with the mounting epoxy, and in parallel with the flexures of
		the sample carrier.  $\sigma_\text{plastic}$ is the stress at which plastic deformation onset.  Mount style A refers to a configuration like in Fig.~1c, where force is applied
		dominantly through the sample's end faces, and B to a configuration like Fig.~1d, where the sample ends are deeply embedded in epoxy such that force is applied mostly through the
		side faces. For all samples, the field axis is the $[001]$ direction.}
	\begin{tabular*}{0.67\columnwidth}{@{\extracolsep{\fill}}cccccc}
		\hline\hline
		& cross & & & & \\
		sample & section & stress & $k$ & & mount \\
		\# & (mm$^2$) & axis & (N/$\upmu$m) & $\sigma_\text{plastic}$ & style \\
		\hline
		1 & $0.18\times0.18$ & $[001]$ & 1.09 & -0.19 GPa at 300 K & A \\
		2 & $0.40\times0.09$ & $[001]$ & 1.09 & -0.20 GPa at 300 K & B \\
		3 & $0.20\times0.20$ & $[001]$ & 1.20 & -0.20 GPa at 100 K & A \\
		4 & $0.30\times0.14$ & $[100]$ & 0.97 & -0.18 GPa at 100 K & B\\
	\end{tabular*}
\end{table}

	\section{Results}
	
	We begin with magnetometry measurements on unstressed samples, to investigate possible spontaneous ferromagnetism. As noted in the Introduction, the cubic symmetry of the lattice
	competes with nearest-neighbour antiferromagnetic interaction, and is expected to cause a small spin rotation out of the plane~\cite{Szunyogh09_PRB, LeBlanc13_PRB}. For the domain
	illustrated in Fig.~\ref{fig:magStructure} (a), the resulting net moment would point along the $[111]$ or $[\bar{1}\bar{1}\bar{1}]$ direction. In Mn$_3$Ir, a net moment of
	7~m$\upmu_\text{B}$/Mn has been predicted~\cite{Chen14_PRL}.  	

	\begin{figure*}[htb]
	\centering
	\includegraphics[width=136mm]{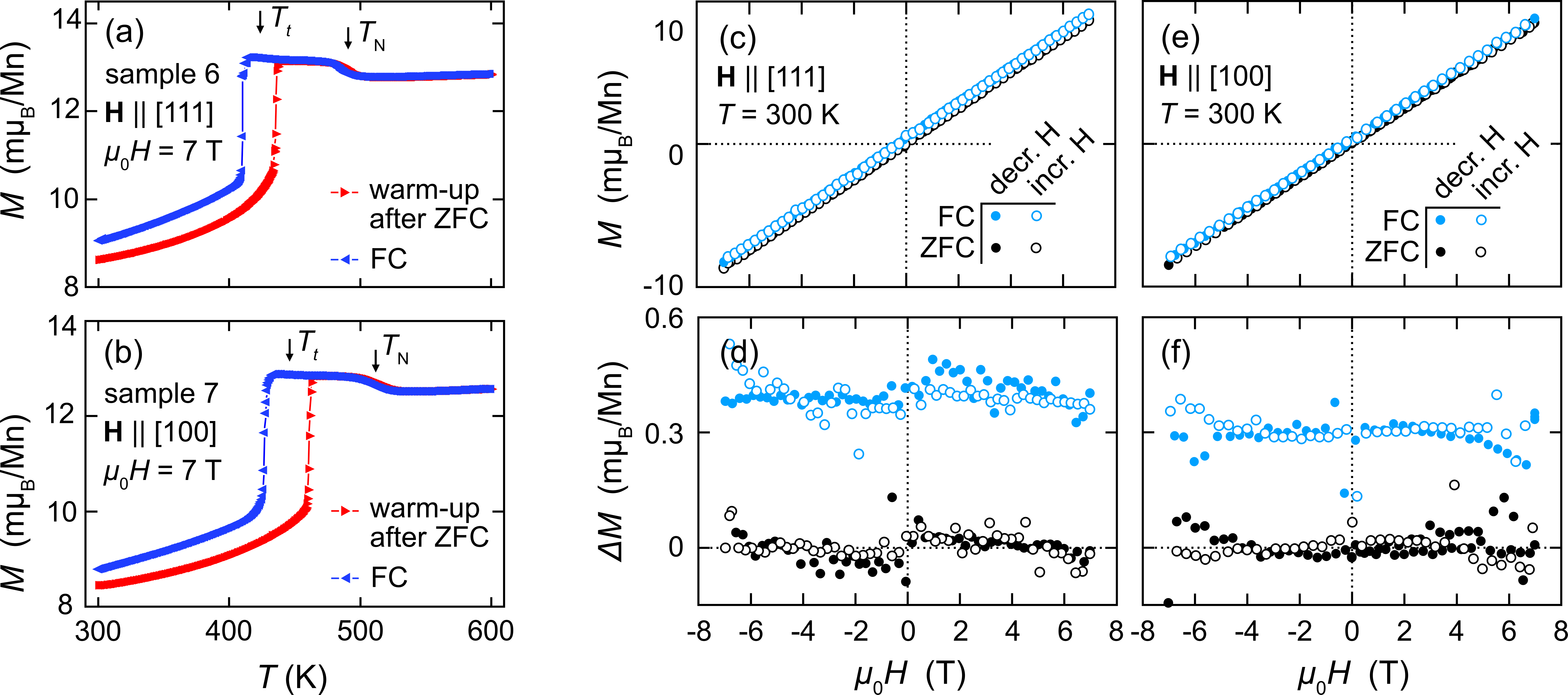}
	\caption{\label{fig:magnetometry} Magnetometry data on unstressed Mn$_3$Pt, setting an upper limit on the spontaneous magnetisation. (a) Magnetisation $M$ against temperature for sample 6 and $\mathbf{H}$ applied along $[111]$. ``ZFC'' indicates zero-field
		cooled, ``FC'' field-cooled, under a 7~T applied field. (b) $M(T)$ for sample~7 and field along $[100]$. (c) $M(H)$ at 300~K for sample 6, after cooling from 600~K in either zero
		field or 7~T. Open points indicate increasing $H$, filled points decreasing $H$. (d) The data from (c), but with a linear fit to the ZFC data subtracted. (e--f) Equivalent to
		panels (c--d) for sample~7 and $\mathbf{H} \parallel [100]$.}
	\end{figure*}
	
	Mn$_3$Pt does not transition directly into the structure illustrated in Fig.~\ref{fig:magStructure} (a), but rather passes through an intermediate phase between $T_\text{t} \approx
	400$~K and $T_\text{N} \approx 475$~K with a collinear spin structure~\cite{Kren66_PhysLett, Kren67_JAP, Kren68_PR}. Data for $\mathbf{H} \parallel [111]$ are shown in
	Fig.~\ref{fig:magnetometry} (a), (c--d), and for $\mathbf{H} \parallel [100]$ in Fig.~\ref{fig:magnetometry} (b), (e--f). The samples were thermally cycled to 600~K, well above $T_\text{N}$, and
	cooled in either 0 or 7~T.  For both samples, a net polarisation of $0.3$--$0.4$~m$\upmu_\text{B}$/Mn is visible below $T_\text{t}$. This is a small net moment, and because it may
	include contributions of polarisation of spins in the vicinity of defects and magnetic domain walls, it should be understood as an upper limit on the intrinsic net moment.
	
\begin{figure*}[htb]
		\centering
		\includegraphics[width=0.9\textwidth]{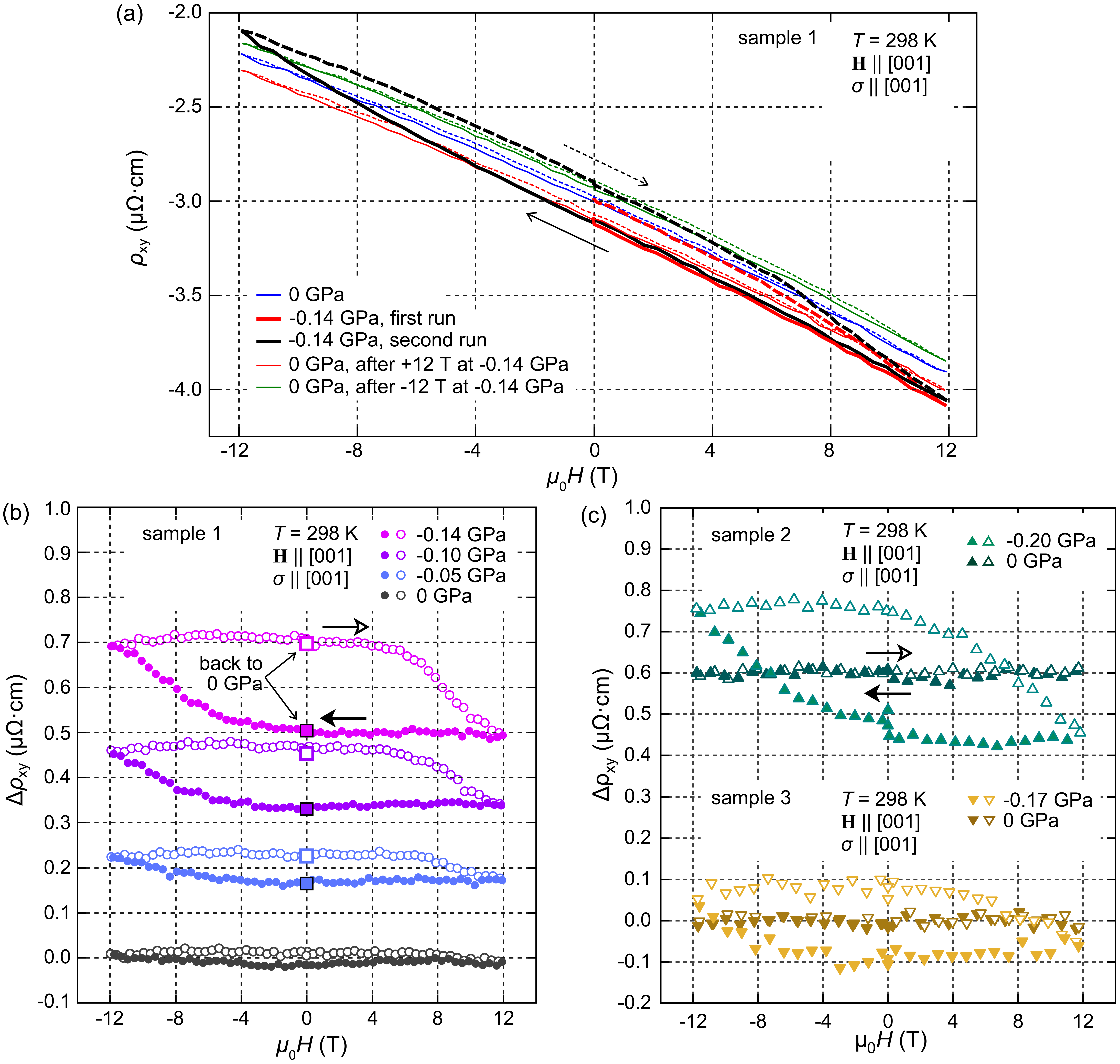}
		\caption{\label{fig:mainResults} (a) Example raw data for sample 1. (b) Background-subtracted data for sample 1. Open points are for increasing $H$, filled points decreasing $H$.
			Under uniaxial stress, the hysteresis in $\rho_\text{xy}(H)$ expands. The large squares are $\rho_\text{xy}$ measured after the applied stress was ramped back to 0~GPa. For
			clarity, data at the separate stresses are offset by 0.2 $\upmu\Omega\cdot$cm. (c) Background-subtracted data for samples 2 and 3. Data from sample 2 are offset by
			1~$\upmu\Omega\cdot$cm.}
	\end{figure*}

	We now discuss measurements of the Hall effect. $\rho_\text{xy}$ data from sample 1 are shown in Fig.~\ref{fig:mainResults} (a).
		
\noindent At zero stress (the blue curve), the field dependence
	of $\rho_\text{xy}$ is dominated by the normal Hall effect. There is resolvable hysteresis, but it is small. Under an applied stress of $-0.14$~GPa (thick red and black
	curves), the hysteresis became much larger. When the stress was released after application of a field of $+12$~T at $-0.14$~GPa, the hysteresis loop closed up again, showing that
	the enhanced hysteresis was not a product of any non-elastic deformation. However, the new curve (the thin red curve) was offset from the original blue curve.  Similarly, the
	zero-stress data recorded after application of $-12$~T at $-0.14$~GPa (the green curve) also showed small hysteresis, but were offset from the original curve in the other
	direction. Evidently, the hysteretic part of $\rho_\text{xy}$ remains at least mostly locked in after the stress is released.
	
	Fig.~\ref{fig:mainResults} (b) shows hysteresis loops with backgrounds subtracted; the background is taken as a quadratic function of $\upmu_0 H$, obtained independently at each
	stress. Each hysteresis loop is recorded following this procedure: (1) under applied stress, $B$ is ramped from $0$ to $+12$~T, and then back to $0$; (2) the stress is then
	released, to measure the Hall effect at zero applied stress; (3) the stress is re-applied, and $B$ ramped from $0$ to $-12$~T, then to $0$; (4) the stress is released again and the
	Hall effect re-measured.  The 0~GPa readings are indicated by large squares: it is apparent that releasing the stress causes essentially no change in $\rho_\text{xy}$.  Hysteresis
	loops for samples 2 and 3, though without the ramps back to 0~GPa, are shown in Fig.~\ref{fig:mainResults} (c). 
	
\begin{figure*}[ptb]
		\includegraphics[width=0.48\textwidth]{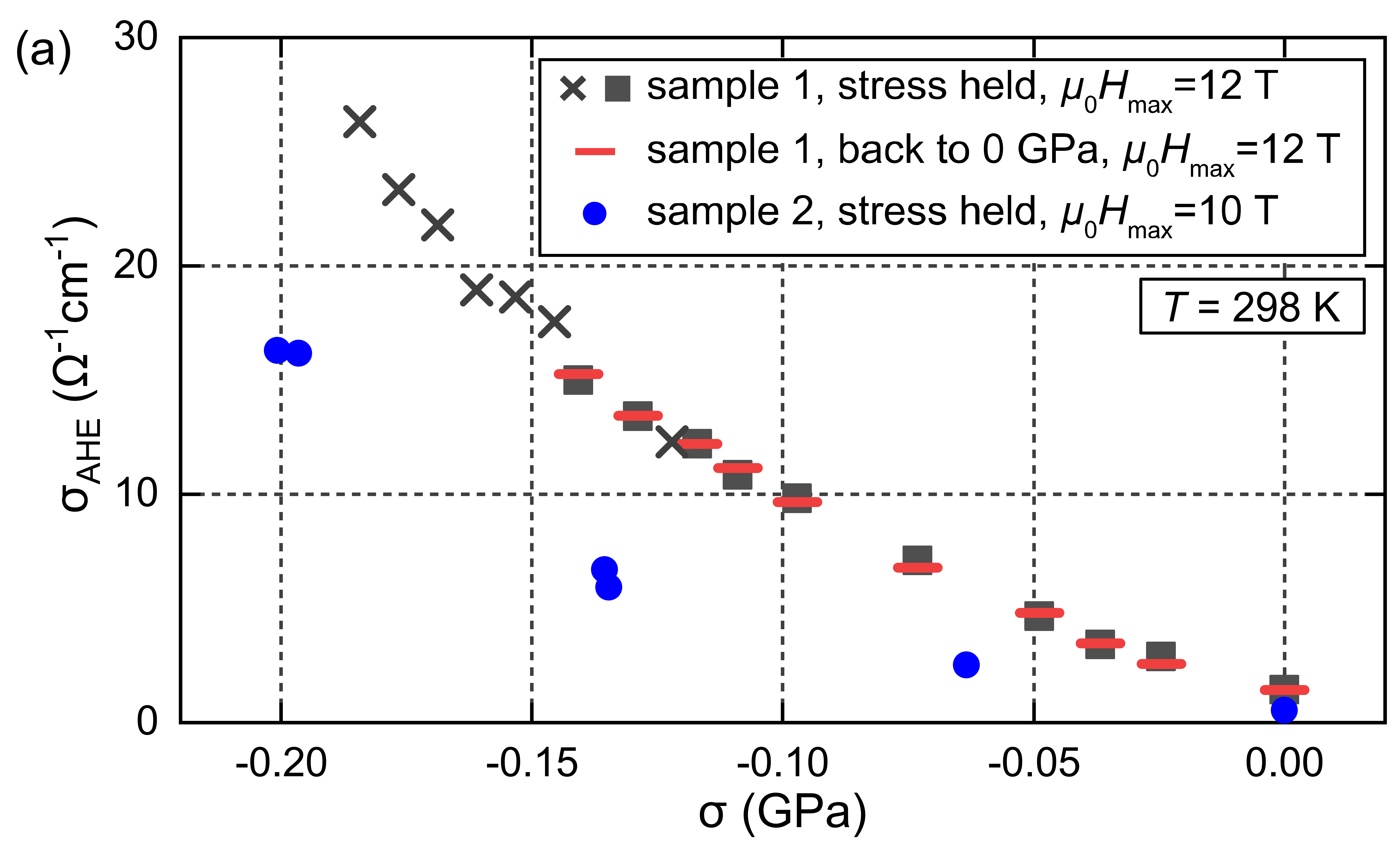} \quad
		\includegraphics[width=0.48\textwidth]{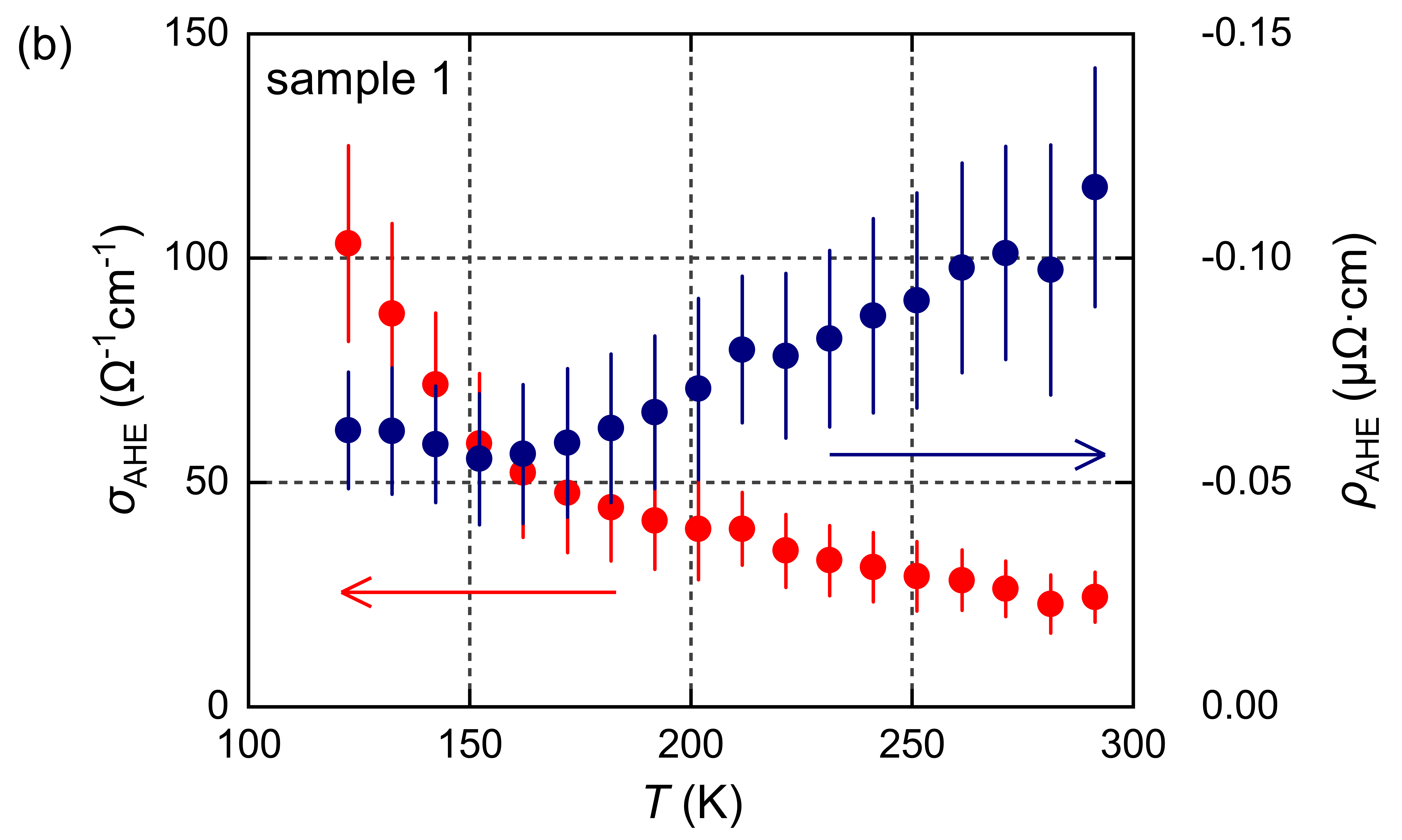}
		\caption{\label{fig:AHE} (a) The anomalous Hall conductivity $\sigma_\text{AHE} = -\rho_\text{AHE}/\rho^2$ of samples 1 and 2, obtained from the stress-induced
			hysteresis loops as described in the text. ``Stress held'' indicates data taken under the indicated applied stress, while ``back to 0 GPa'' indicates data taken when the stress was
			released, but after application of the indicated stress and $H = \pm H_\text{max}$. For sample 1, the points indicated by the gray X's were taken over a year after the points
			indicated by the gray squares. The fact that $|\sigma_\text{AHE}|$ continues to increase as the sample is compressed shows that within the stress and field limits of this
			measurement we were not able to fully align the magnetic domains. (b) $\sigma_\text{AHE}$ from sample 1 after application of $\sigma = -0.14$~GPa, $\upmu_0H = \pm 12$~T at 298~K,
			as a function of temperature.}
	\end{figure*}

	The anomalous Hall conductivity $\sigma_\text{AHE} = -\rho_\text{AHE}/\rho^2$ for samples 1 and 2 is shown in Fig.~\ref{fig:AHE} (a). $\rho_\text{AHE}$ at each stress is
	taken as $\frac{1}{2}[\rho_\text{xy}^+(H=0) - \rho_\text{xy}^-(H=0)]$, where $\rho_\text{xy}^\pm$ is the Hall resistivity after application of $\upmu_0 H = \pm 12$~T for sample~1 and
	$\pm 10$~T for sample~2. $\rho_\text{xx}$ is taken from sample~4, where the contact configuration permitted accurate measurement; see Fig.~\ref{fig:resistivity}. The gray squares
	in Fig.~\ref{fig:AHE} (a) show $\sigma_\text{xy}$ derived from data taken while holding the applied stress, and the red lines after the stress was ramped back to zero. There is no
	significant difference between them, which shows that any stress-induced ferromagnetic moment contributes negligibly to the AHE. 
	
	After a year-long pause, we attempted to achieve saturation of the AHE in sample~1 by driving it to higher compression. These data points are indicated by the X's in
	Fig.~\ref{fig:AHE} (a). Their position is shifted by hand along the stress axis to match the earlier data set, compensating for relaxation in the mounting epoxy that occurred during
	the long pause. These data make clear that within the limits of our maximum field (12 T) and the elastic limit of the samples ($\sim 0.2$~GPa) clear saturation could not be
	achieved, and it is not clear how much higher field and/or stress would be required. The coercive field for strained bulk Mn$_3$Pt is evidently far higher than that observed in
	the thin film study, where it was $\sim 0.1$~T~\cite{Liu18_NatureElectronics}. Disorder and/or interface effects may have suppressed the coercive field in the thin films.
	
	Separately, we note that in Ref.~\cite{Liu18_NatureElectronics} the intrinsic AHE was calculated to have a strain dependence. According to this calculation, at the largest stress
	that we reached the anomalous Hall conductivity would be 4\% larger than at zero stress. Such a change would be at the edge of our present sensitivity.
	
	Fig.~\ref{fig:AHE} (b) shows $\sigma_\text{AHE}$ of sample 1, after training the magnetic structure at 300~K with $\sigma = -0.14$~GPa and $\upmu_0 H = \pm 12$~T, as a function of
	temperature. $\sigma_\text{xy}$ increases steadily with cooling, as was observed with Mn$_3$Ge~\cite{Kiyohara16_PRA}. 
	
\begin{figure}[ptb]
		\centering
		\includegraphics[width=96mm]{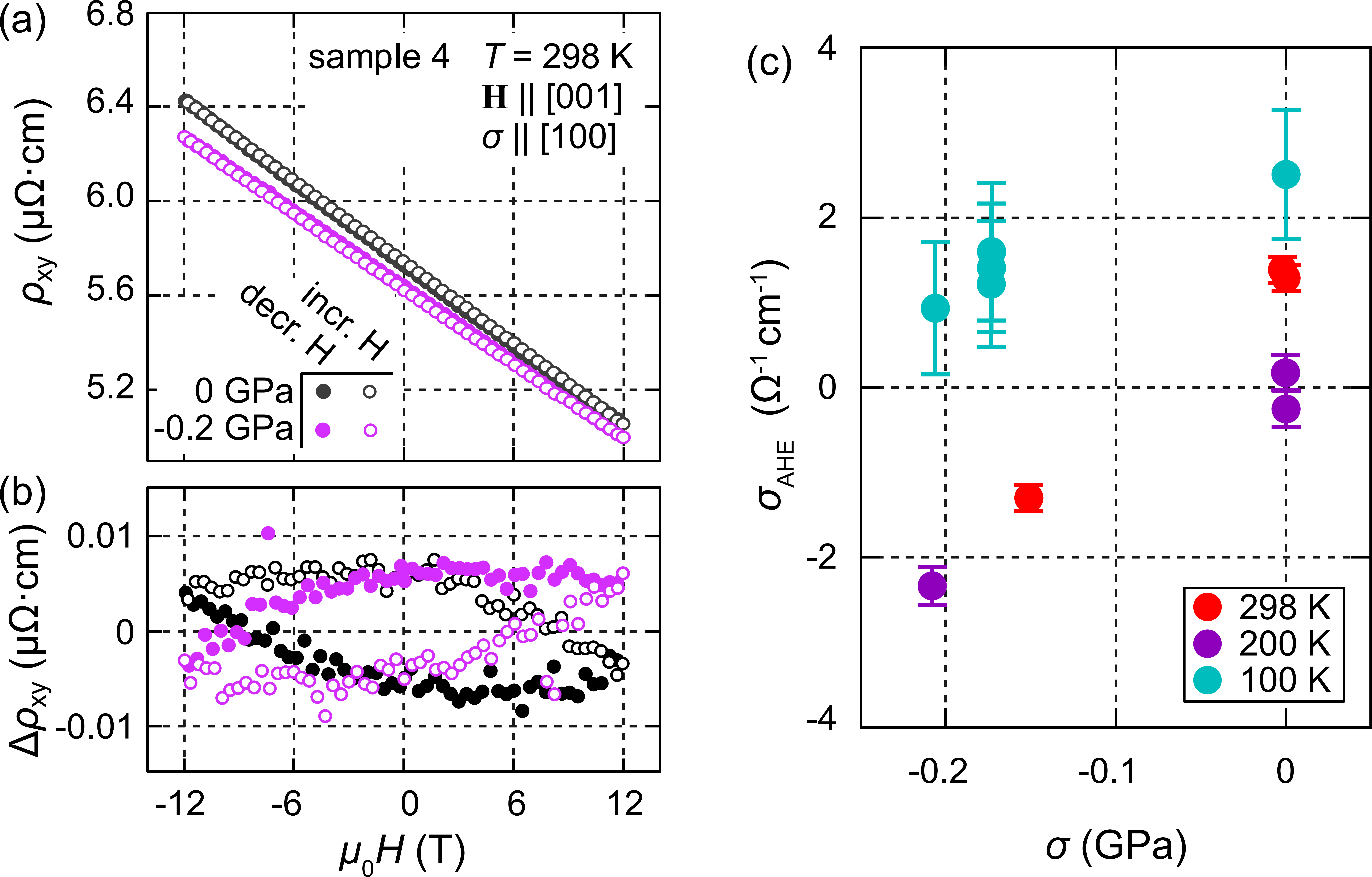}
		\caption{\label{fig:alternativeConfiguration} (a) $\rho_\text{xy}(H)$ for sample 4, where field was applied along $[001]$ and stress along $[100]$. Open/closed symbols
			indicate increasing/decreasing $H$, respectively. (b) Data from (a) with quadratic backgrounds subtracted. (c) Anomalous Hall conductivity for this sample at
			selected temperatures and stresses.}
	\end{figure}

	We now show, in Fig.~\ref{fig:alternativeConfiguration}, stress-induced hysteresis for sample 4, which was the one with a transverse configuration: compression along $[100]$,
	$\mathbf{H} \parallel [001]$. As for sample~1, for sample~4 there was discernible though small hysteresis at zero applied stress. Measurements at three different
	temperatures (100, 200, and 298~K), shown in Fig.~\ref{fig:alternativeConfiguration} (c), reveal a consistent effect of the applied stress: $\sigma_\text{AHE}$ decreases under
	compression, a change of the opposite sign to that observed when $\sigma \parallel [001]$. However, the quantitative effect is much smaller than for $\sigma \parallel [001]$.

	\section{Discussion and Conclusion}
	
	We have shown that the anomalous Hall effect of Mn$_3$Pt can be trained by simultaneous application of uniaxial stress and field. In other words, uniaxial stress reduces the
	coercive field for reorientation of the antiferromagnetic domains. Stress that is longitudinal with respect to the field is more effective than transverse stress. We discuss
	below three possible mechanisms by which uniaxial stress might reduce the coercive field. (1) piezomagnetism due to out-of-plane rotation of the spins (2) piezomagnetism due
	to in-plane rotation of the spins; and (3) approach to an alternative magnetic ground state. 
	
	Due to concerns about plastic deformation we did not attempt to measure the piezomagnetism of Mn$_3$Pt. In the epitaxial thin film study of
	Ref.~\cite{Liu18_NatureElectronics}, a net magnetic moment of 4.6~m$\upmu_\text{B}$/Mn was obtained for a 20~nm-thick film with $c/a = 0.990$. For bulk samples, that ratio of $c/a$
	would be obtained under a uniaxial stress of $\sigma \approx -0.8$~GPa~\footnote{The bulk modulus of Mn$_3$Pt is $115 \pm 8$~GPa~\cite{Yasui87_JPSJ}, and the Poisson's ratio is
		0.35~\cite{Liu18_NatureElectronics}, which together yield a Young's modulus of $106 \pm 7$~GPa}, indicating a piezomagnetic coefficient of $\approx 6$~(m$\upmu_\text{B}$/Mn)/GPa.  At
	our largest stress of $\sigma \sim -0.2$~GPa, a piezomagnetic moment of 1.2~m$\upmu_\text{B}$/Mn would be obtained, well above our upper limit for the spontaneous magnetisation in
	unstressed Mn$_3$Pt. For comparison, 8~(m$\upmu_\text{B}$/Mn)/GPa has been measured for in-plane stress applied to Mn$_3$Sn~\cite{Ikhlas22_NatPhys}, and $\sim
	100$~(m$\upmu_{B}$/Mn)/GPa has been reported for Mn$_3$NiN~\cite{Boldrin18_ACS}.
	
	In the triangular state at $T < T_\text{t}$, there are eight equivalent domain types. There are four possible spin planes: $(111)$, $(1\bar{1}1)$, $(\bar{1}11)$, and
	$(\bar{1}\bar{1}1)$, and for each there are two domain types, related by time reversal. Following Ref.~\cite{Chen14_PRL}, these can be classed as A-phases, where the spins point
	outward with respect to the triangle that is higher along $[001]$, and B-phases, where they point inward--- see Fig.~\ref{fig:magStructure} for an illustration. In the
	experimental geometry here, where our Hall effect measurements probe effective fields parallel to $[001]$, the AHE from A and B phases will be opposite.
	
	Piezomagnetism is given by $M_i = \sum\limits_{j,k}\Lambda_{ijk}\sigma_{jk}$, where $\mathbf{M}$ is the induced magnetic moment, $\sigma$ is stress, and $\Lambda_{ijk}$ is the
	piezomagnetic tensor. For the magnetic space group of Mn$_3$Pt, R$\bar{3}$m$^\prime$, $\Lambda_{zzz}$ and $\Lambda_{zxx}$ are both nonzero, though with different coefficients,
	meaning that both $z$-axis and $x$-axis uniaxial stress can induce a $z$-axis magnetic moment. One possibility is that the main effect of the applied uniaxial stress is to alter
	the out-of-plane spin canting. However, this form of piezomagnetism would not account for the differing effects of longitudinal versus transverse stress, because for each of the
	possible spin planes the geometry is equivalent for compression along $[001]$ and $[100]$, and the induced moment along $[001]$ would be the same.
	
	In-plane spin rotation could explain the difference between longitudinal and transverse stress. The black arrows in Fig.~\ref{fig:magStructure} indicate the expected spin
	rotation under compression along [001] for the A and B phases, under an assumption of in-plane spin rotation. Spin 1 does not rotate because there is no symmetry breaking between
	clockwise and counter-clockwise rotation. Spins 2 and 3 rotate away from spin 1 because the bond length to spin 1 is reduced. That the magnetic interaction energy increases with
	reduced bond length is shown by the fact that $T_\text{N}$ under strong hydrostatic compression, where the collinear phase is suppressed and the transition at $T_\text{N}$ is
	directly into the triangular phase, increases with compression~\cite{Yasui87_JPSJ}. Measurements of spin waves show in addition that nearest-neighbor interaction is
	dominant~\cite{Ikeda03_JPSJ}. For the A phase, the resulting net moment points along $[11\bar{2}]$. Compression along $[100]$, on the other hand, would cause spins 1 and 3 to
	rotate, resulting in a net moment for the A phase along $[\bar{1}11]$. Its projection along $[001]$ is opposite to that induced by $[001]$ compression, and half as large. The same
	will be true for the other possible spin planes. Therefore, for a given field along $[001]$, opposite domain types --- A versus B --- should be selected by longitudinal versus
	transverse stress, with transverse stress being less effective. We note that orbital polarisation could add comparably to, and even reverse, the net moment from spin
	rotation~\cite{Sandratskii96_PRL, Chen20_PRB}, but this anisotropy remains.
	
	There is a strong possibility of a more complicated response to the applied stress. Stress along $\langle 100 \rangle$ directions is not aligned with the principal axes of
	the magnetic order or the local easy axes of the Mn spins, making it likely that the applied stress will eventually favour an alternative ground state. In other magnetic systems,
	stresses well below 1~GPa have been observed to qualitatively change the magnetic structure~\cite{Park18_PRB, Sun21_NJP}. There is evidence for
	closely-spaced ground states in Mn-based systems. In Mn$_3$GaN, a state where the spins are rotated by $90^\circ$ with respect to the Mn$_3$Pt structure may be the ground
	state~\cite{Lukashev08_PRB, Zemen17_PRB, Samathrakis20_PRB}, and may be close to the ground state in Mn$_3$Pt~\cite{Bai21_PRB}. In Mn$_3$Ge and Mn$_3$Ga there is a strong
	susceptibility for rotation of the spins out of the plane~\cite{Sukhanov18_PRB, Song21_APL}. Therefore, it is
	possible that approach to an alternative ground state, and associated weakening of the magnetic anisotropy, contributes substantially to the observed strain-driven suppression of
	the coercive field.
	
	As in the thin film study, the AHE observed here is of the same order of magnitude as predicted theoretically~\cite{Zhang17_PRB, Liu18_NatureElectronics}. Without demonstration
	of saturation of the magnetic domains, we cannot offer detailed comments on the magnitude of $\rho_\text{AHE}$.
	
	In conclusion, we have shown that uniaxial stress reduces the coercive field for training the magnetic structure of Mn$_3$Pt, allowing experimental observation, for the first
	time, of the anomalous Hall effect in bulk samples of a cubic member of the Mn$_3$X family. The anomalous Hall effect induced by stress and field remains locked in after the stress is released, which shows that the effect of any
	stress-induced ferromagnetic moment on the Hall effect is negligible. Longitudinal stress reduces the coercive field faster than transverse stress, an observation suggesting
	that in-plane spin rotation is the dominant response to the applied stress. This hypothesis is not the only possibility, and so should be tested through direct observation of the
	piezomagnetism. Saturation of the AHE may prove possible through combinations of stress and field directions not probed here, allowing precise comparison with calculations on
	single crystals, and we therefore encourage further experimentation.
	
	\section*{Acknowledgements}
	
	We thank Peter Milde, Yan Sun, and Binghai Yan for useful discussions. We thank in addition Tomoya Higo and Satoru Nakatsuji for critical readings of our manuscript, and Hua
	Chen for assistance with the piezomagnetic tensor. We also thank Renate Hempel-Weber and Ulrike Lie\ss{} for technical
	assistance. We acknowledge the financial support of the Max Planck Society. BZ, CF, APM, and CWH acknowledge in addition financial support from the the Deutsche Forschungsgemeinschaft
	through SFB 1143 (Project ID 247310070). KM and CF acknowledge financial support from the European Research Council (ERC), Advanced Grant No. 742068 ‘TOPMAT’; European Union’s
	Horizon 2020 research and innovation program (Grant Nos. 824123 and 766566). Additionally, KM acknowledges funding support from Max Planck Society under Max Planck-India partner
	group project and the Board of Research in Nuclear Sciences (58/20/03/2021- BRNS, DAE-YSRA), Department of Atomic Energy (DAE), Government of India. H.M.L.N. acknowledges support from the Alexander von Humboldt Foundation through a Research Fellowship for Postdoctoral Researchers.
	
	\section*{References}
	\bibliography{draft_arxiv}

\end{document}